\documentclass{article}

\title{Motion in random fields - an application to stock market data}
\author{ James P. Gleeson\\
\emph{Department of Applied Mathematics,}\\
\emph{University College Cork, Ireland.}\\
 Email: j.gleeson@ucc.ie}
\date{November 7, 2003.}
\usepackage{amsmath}
\usepackage{epsfig}

\def\t{\tau}
\def\p{\partial}   

\def\a{\alpha}
\def\b{\beta}
\def\w{\omega}
\def\eps{\epsilon}

\begin{document}
\maketitle
\begin{abstract}
A new model for stock price fluctuations is proposed, based upon
an analogy with the motion of tracers in Gaussian random fields,
 as used in turbulent dispersion models and in studies of
transport in dynamically disordered media. Analytical and
numerical results for this model in a special limiting case of a
single-scale field show characteristics similar to those found in
empirical studies of stock market data. Specifically, short-term
returns have a non-Gaussian distribution, with super-diffusive
volatility, and a fast-decaying correlation function. The
correlation function of the absolute value of returns decays as a
power-law, and the returns distribution converges towards Gaussian
over long times. Some important characteristics of empirical data
are not, however, reproduced by the model, notably the scaling of
tails of the cumulative distribution function of returns.
\end{abstract}
\section{Introduction}
Random fluctuations in stock market prices  have long fascinated
investors and mathematical modelers alike. Although the investors'
hopes of accurately predicting tomorrow's share price appear to be
in vain, models which limit themselves to statistical
characteristics such as distributions and correlations have had
some success. Bachelier's classical model \cite{Bachelier} treats
the stock price $S(t)$ as a random walk, leading to the conclusion
that the distribution of prices is Gaussian. Samuelson
\cite{Osborne} instead describes  the log-price
\begin{equation}
x(t)=\ln[S(t)/S(0)]
\end{equation}
 as a random walk, and
therefore concludes that the stock price $S(t)$ should have a
log-normal distribution. This model remains in common usage,
despite the shortcomings listed below, not least because it
permits the derivation of an equation for pricing options and
other financial derivatives, e.g., the famous Black-Scholes
equation \cite{Hull}. Nevertheless, empirical evidence from stock
market data indicates that the random walk model inadequately
describes many important features of the stock price process. The
following stylized facts are accepted as established by these
studies \cite{Hull,Campbell}:
\begin{itemize}
\item[(i)] Short-term returns are non-Gaussian, with `fat tails'
and high central peaks. The center of the returns distribution is
well fitted by L\'{e}vy distributions \cite{Nat95}. Recent studies
indicate that the tails of the returns distribution decay as
power-laws \cite{Gop}. As the lag time increases, the returns
distributions slowly converge towards Gaussian; recent analysis of
the Standard \& Poor 500 index (S\&P500) estimates this convergence is seen only on lags
longer than 4 days \cite{Gop}.
 \item[(ii)] The correlation function of returns decays
 exponentially over a short timescale, consistent with market
 efficiency. However, the correlation function of the
 \emph{absolute value} of the returns shows a much slower
 (power-law) decay \cite{Gop}.
\item[(iii)] The volatility (standard deviation of returns) grows
like that of a diffusion (random walk) process, i.e. as the square root
of the lag time, for lag times longer than about 10 minutes.
However, higher frequency (shorter lag) returns demonstrate a super-diffusive
volatility, which can be fitted to power-laws with
exponents found to range between 0.67 and 0.77
\cite{Gop,Masolivier00}. \item[(iv)] The distribution of stock
price returns exhibits a simple scaling: in \cite{Nat95} a
power-law scaling of the peak of the returns distribution $P(0)$
with lag time is shown to hold across many magnitudes of lag
times. The exponent of the power-law is approximately $-0.7$.
\end{itemize}
Most of the references cited here examine data from the Standard \& Poor 500 index (S\&P500), but other international markets are found to behave similarly \cite{Gop}.

The predictions of the random-walk model can be investigated using
the random differential equation
\begin{equation}
\frac{d x}{d t}=u_w(t), \label{1red}
\end{equation}
which yields the log price $x(t)$ from each realization of the
random function $u_w(t)$. The classical random walk follows from
taking $u_w(t)$ to be a Gaussian white-noise  process with zero
mean and (auto)correlation function \cite{VK}:
\begin{equation}
\left< u_w(t)u_w(t^\prime)\right>=\a^2 \delta(t-t^\prime),\label{2red}
\end{equation}
where 
$\delta(t)$ is the Dirac delta function, and the angle brackets
denote averaging over time or over an ensemble of realizations.
White noise is the formal derivative of a Wiener process, which
connects (\ref{1red}) to rigorous methods for stochastic
differential equations \cite{Arnold}. Note that throughout this
paper all stochastic processes are assumed (unless explicitly
stated otherwise) to have zero mean, zero drift, and to be
statistically stationary.

Equation (\ref{1red}) is easily solved, and leads to results which
do not agree with the empirical facts (i)-(iv) listed above.
Defining the return on the stock price $S(t)$ at time $t$ as the
forward change in the logarithm of $S(t)$ over the lag time
$\Delta$:
\begin{eqnarray}
r_\Delta(t)&=&x(t+\Delta)-x(t)\nonumber\\
&=&\ln S(t+\Delta)-\ln S(t) ,
\end{eqnarray}
it is found that the white-noise case leads to probability
distribution functions (PDFs) for $r_\Delta$ which are Gaussian at
all lags $\Delta$, in contrast to (i). The lack of memory effects
in white-noise leads to correlation functions which decay
immediately to zero, unlike (ii). The volatility grows diffusively
for all lag values, and so has no super-diffusive range, while the
scaling of $P(0)$ follows a power-law different from that found
for actual prices, see (iv) above.

A natural generalization of the white-noise model again takes a
form similar to (\ref{1red})
\begin{equation}
\frac{d x}{d t}=u_c(t),\label{4red}
\end{equation}
but with $u_c(t)$ now being a non-white or colored process,
incorporating some memory effects. For the particularly simple
case of Gaussian noise, it can be described fully by its
correlation function
\begin{equation}
\left< u_c(t) u_c(t^\prime)\right> = \a^2 R(t-t^\prime) \label{5red}.
\end{equation}
Here $R(t)$ decays from a maximum of unity at $t=0$ to zero as
$t\rightarrow \infty$, and $\a^2$ is the variance of the process $u_c$. Gaussian noise is attractive because it
appears naturally as the limiting sum of many independent noise
sources under the central limit theorem \cite{VK}, and because of
its analytical tractability. Indeed equation (\ref{4red}) is again
easily solvable for Gaussian $u_c$, but leads once more to
Gaussian-distributed returns at all lag times, in contrast to (i).
Moreover, the correlation of absolute returns does not decay more
slowly that the returns correlation, as required by (ii).

These failings of simple differential equation models have
inspired many attempts to reproduce the empirical facts (i)-(iv)
using various stochastic models. As a sample of just a few of
these, we list the truncated L\'{e}vy flights model
\cite{Mandelbrot,MS94}, ARCH and GARCH models \cite{Engle,
Bollerslev}, non-Gaussian Ornstein-Uhlenbeck processes
\cite{BNShepherd}, and a model based on a continuous superposition
of jump processes \cite{MasolivierPRE, Masolivier00}. The first
three of these were reviewed in \cite{MS98}. An ideal model should
reproduce all the experimental facts (i)-(iv), by giving a simple
picture of the stochastic process underlying the stock price time
series, but no model has yet been found which fulfills all these
criteria \cite{MS99}.

In this paper we examine the results of generalizing the random walk models (\ref{1red}) and (\ref{4red}) to motion in a random field:
\begin{equation}
\frac{d x}{d t}=u(x,t). \label{6red}
\end{equation}
Here $u(x,t)$ is a Gaussian random field with zero mean, which is
fully described by its correlation function
\begin{eqnarray}
\left< u(x,t)u(x^\prime,t^\prime)\right>&=&\a^2 Q(x-x^\prime,t-t^\prime) \label{7red}\\
&=& \a^2 S(x-x^\prime)R(t-t^\prime) \label{8red}.
\end{eqnarray}
For simplicity we take the field to be homogeneous and stationary;
the factorization in (\ref{8red}) of the correlation into separate
time- and $x$-correlations is for ease of exposition only, and
more complicated inter-dependent correlations can be considered.
The averaging procedure in equation (\ref{7red}) is over an
ensemble of realizations of the random field $u$, or using a
uniform measure over all possible $(x,t)$ values.

Equation (\ref{6red}) implies that the log-price $x(t)$ changes in
a random fashion, but the rate of change is randomly dependent on
both time and the current value of $x$. Note that random-walk
models (\ref{1red}) and (\ref{4red}) may be recovered by taking
$S(x)\equiv 1$ and choosing the time correlation $R(t)$
appropriately. We will show that the $x$-dependence in the noise
term yields qualitatively different results to the random walk
models, and that many of the empirical observations (i)-(iv) of
stock markets may be reproduced by (\ref{6red}).

The $x$-dependent noise term in (\ref{6red}) is motivated by
recent studies in turbulent dispersion \cite{Flatness} and
transport in dynamically disordered media \cite{Witkoskie}. These
investigations employ equation (\ref{6red}), but with a rather
different interpretation: $x(t)$ is the position vector (in 2 or 3
dimensions) of a passive tracer, and $u(x,t)$ is the velocity
vector field which transports the tracer. The tracer is called
\emph{passive} because it is assumed not to affect the velocity
field by its presence, and so $u(x,t)$ is a prescribed random field. An
instructive example is the motion of a small buoy on the surface
of the ocean \cite{Balk}: the two-dimensional vector $x(t)$ then
describes the position of the buoy (by its longitude and latitude,
say) while the tracer is moved by the ocean waves according to
(\ref{6red}), with the Gaussian field $u(x,t)$ providing an
approximate description of the ocean wave field.

The distribution of such tracers resulting from motion in a
Gaussian field is known to be non-Gaussian, at least over short to
intermediate timescales \cite{Flatness, Kraichnan}. Crucial to
understanding this effect is the difference between the
\emph{Eulerian} velocity $u(x,t)$, i.e.,  the velocity measured at
the fixed location $x$ at time $t$, and the \emph{Lagrangian}
velocity
\begin{equation}
v(t)=u(x(t),t) \label{9red},
\end{equation}
which is the time series of velocity measurements made by the
tracer itself as it moves through the random field \cite{McComb}.
When there is no $x$-dependence in the velocity field $u$, the
Eulerian and Lagrangian velocities coincide, but otherwise a
Gaussian Eulerian velocity field can yield non-Gaussian Lagrangian
processes. The relationship between Eulerian and Lagrangian random processes is an active research area, especially in the case of a compressible (non-zero divergence) velocity field which is of interest here \cite{Zirbel, Vlad, JPGComment}.

Our model of the log-price uses a scalar ``position'' $x(t)$ and
``velocity'' $u(x,t)$ rather than the vector-valued analogues in
the turbulent dispersion problems described above. Nevertheless,
the concept of Eulerian and Lagrangian velocities carries over to
the one-dimensional case, and so we will refer to $x(t)$ as the
``price tracer'', with its ``velocity'' $u(x,t)$ related by
equation (\ref{6red}). The fact that the Eulerian velocity field
$u(x,t)$ is Gaussian allows us to obtain analytical results, and
may be justified as a consequence of the central limit theorem
applied to the sum of many independent forces on stock prices. In
this paper we will not further pursue a microeconomic
justification \cite{Lux,Appleby} for (\ref{6red}), but rather
concentrate on demonstrating that it may yield some of the
observed statistical properties of empirical stock market data.

The remainder of this paper is structured as follows. In section 2
the consequences of the $x$-dependence of the noise in equation
(\ref{6red}) are examined through some simple examples, and a
special limiting case (single-scale field) is highlighted. The
statistics of the Lagrangian velocity (\ref{9red}) in the
single-scale case are found analytically in sections 3 and 4, and
are related to the statistics of the returns $r_\Delta$. Numerical
simulations of equation (\ref{6red}) are considered in section 5,
and a simple method for generating Monte-Carlo time series in the
single-scale case is found. Results from such simulations are
presented in section 6, and compared to the stylized facts
(i)-(iv), as presented in \cite{Gop} and \cite{Nat95}. Section
7 comprises of discussion of the results and directions for future
work.

\section{X-dependent velocity fields}

The concept of $x$-dependence in the velocity of the log-price
$x(t)$ was introduce in section 1. To gain some intuitive
understanding of the effect of such $x$-dependence, we first
return to the random walk model with colored noise $u_c(t)$,
independent of $x$, as in (\ref{4red}) and (\ref{5red}):
\begin{equation}
\frac{d x}{d t}=u_c(t) \text{, with } \left<
u_c(t)u_c(t^\prime)\right> =\a^2 R(t-t^\prime). \label{10red}
\end{equation}
For clarity, we consider in this section a specific form for the time
correlation $R(t)$:
\begin{equation}
R(t)=\exp\left(-\frac{t^2}{2 T^2}\right), \label{Rexp}
\end{equation}
where a characteristic decay time $T$ has been introduced. Figure
1 shows the contours of a single realization of the
$x$-independent $u_c(t)$, negative values of the velocity being
denoted by shaded areas and dotted contour lines. Time increases
along the horizontal axis, with the stock log price $x$ measured
on the vertical axis. The motion of a stock tracer with initial
condition $x(0)=0$ is shown also; note the negative rate of change
of stock price in the shaded regions (negative velocity), and the
positive rate of change in the regions where $u>0$. Because there
is no $x$-dependence in (\ref{10red}), the velocity field depends
only on time, and so the contours are vertical lines. By contrast,
a simple $x$-dependent velocity field of the type proposed in
(\ref{6red}) has a non-trivial $x$-correlation function $S(x)$ in
equation (\ref{8red}); for example in Fig. 2 we use
\begin{equation}
S(x)=\exp\left( -\frac{x^2}{2 l^2}\right) \label{Sfig2},
\end{equation}
where $l$ defines a correlation scale ($l \rightarrow \infty$
recovers the $x$-independent case). The contours in Fig. 2 display
a more complicated structure than in Fig. 1, and this is reflected
in the observed motion of the stock prices. The three parameters
$\a$, $T$ and $l$ characterizing the random field may be combined
into one dimensionless quantity
\begin{equation}
\eps=\frac{l}{\a T}, \label{defeps}
\end{equation}
and if the time, log price and velocity are re-scaled using these
parameters:
\begin{eqnarray}
t &=& T \tilde t \nonumber\\
x &=& l \tilde x \nonumber\\
u &=& \a \tilde u,
\end{eqnarray}
then the equation of motion (\ref{6red})  becomes
\begin{equation}
\eps \frac{d \tilde x}{d \tilde t}=\tilde u(\tilde x,\tilde t).
\label{5resc}
\end{equation}
Note that the re-scaled velocity field $\tilde u$ has unit
variance. In Fig 2. the stock price motions resulting from solving
(\ref{5resc}) are shown for $\eps$  values of $\eps=1$,
$\eps=0.1$, and $\eps=0.01$.
When $\eps$ is large, the length scale $l$ of the randomness in
$x$ is much larger than the effect $\a T$ of the random variation
in time --- this is analogous to the case of weak space dependence
in turbulent dispersion problems studied in \cite{Flatness} using
perturbation theory. In the case of $\eps \gg 1$, small deviations
of the stock price distribution from a Gaussian shape are found,
with kurtosis less than 3. However, the studies of empirical data
mentioned in section 1 indicate that the kurtosis of stock returns
is much \emph{larger} than 3, and so the weakly space-dependent
theory does not seem relevant here.

The opposite limit, i.e., $\eps \ll 1$ appears more promising. In
Fig. 2, we observe that the stock prices for parameter values
$\eps=0.1$ and $\eps=0.01$ display some interesting generic
features. Constrained by the equation of motion to move downwards
in shaded regions (where the velocity $\tilde u$ is negative), and
upwards in unshaded regions, the stock tracers tend to become
trapped near curves of $\tilde u=0$, with $\tilde u$ positive
below them and negative above them. These stable positions
disappear when the zero-velocity manifold `folds over', and the
stock tracer is then driven by  equation (\ref{5resc}) to move
quickly to another slow manifold. A complete understanding of the
dynamics of the system as $\eps\rightarrow 0$  requires a singular
perturbation approach \cite{singpertref} to equation
(\ref{5resc}), but the important feature for our work is the
following of the zero-velocity manifold by the stock tracer,
punctuated by fast jumps.

An analytical approach to the general $\eps \rightarrow 0$ case
has not yet been found, but some interesting features are
highlighted by choosing a particularly simple form for the
$x$-correlation function:
\begin{equation}
S( x)=\cos(k  x) \label{Ssinglesc},
\end{equation}
where $k$ is a constant. 
This correlation function corresponds to a so-called
`single-scale' Gaussian velocity field, which (as shown in
\cite{exact}) can always be written in the form
\begin{equation}
u(x,t)=f(t) \cos(k x)+g(t) \sin(k x), \label{usinglesc}
\end{equation}
where $f(t)$ and $g(t)$ are independent Gaussian random functions
of time, each with zero mean, variance $\a^2$,  and a given
correlation function $R(t)$:
\begin{equation}
\left< f(t)f(t^\prime) \right> =\left< g(t)g(t^\prime)
\right>=\a^2 R(t-t^\prime). \label{deffg}
\end{equation}
An example of motion in such a single-scale field is shown in Fig.
3. Note the periodicity in the vertical direction, and the lack of
jumps. In the remainder of this paper we will concentrate on this
simplified example and show that many of the statistical
characteristics of the resulting stock returns are remarkably
similar to those observed in empirical data.

\section{Lagrangian velocity and PDF}
In this section we consider the statistical characterization of
motion along the $u=0$ curves of the single-scale random field
(\ref{usinglesc}), according to the equation of motion
(\ref{6red}), and with the Gaussian random
functions $f(t)$ and $g(t)$ defined as in (\ref{deffg}). Again,
the analogue with tracer motion on the surface of a turbulence
ocean proves instructive. The Eulerian velocity at a point $x$ at
time $t$ is defined by the random field (\ref{usinglesc}). Each
tracer particle moving through this field will, however,
experience its own time history of velocity variations; this is
called the Lagrangian velocity of the tracer $v(t)$. The crucial
point in our use of random-field models of stock price motion is
that the Eulerian field may, as in (\ref{usinglesc}), have
Gaussian statistics, while the Lagrangian velocity (that is, the
velocity `felt by the tracer') need not be Gaussian. Some
intuitive notion of why this might be may be gained from
considering the fact that the velocity at space-time points
$(x,t)$ picked at random from the Eulerian field has a Gaussian
distribution; on the other hand, the dynamics of the tracer
particles mean that they are more likely to cluster in regions of
low velocity, with high-velocity transitions between, meaning both
zero values of $v$ and high values are more likely than for a
Gaussian distribution. This argument holds for any random field,
but as we show in this section, a quantitative description may be
derived in the case of a single-scale field (\ref{Ssinglesc}) for
the $\eps\rightarrow 0$ limit.

In the limit of small $\eps$, we saw in the previous section that
the stock tracer motion is confined to curves with $u=0$. While
the Eulerian velocity on these curves is zero by definition, the
Lagrangian velocity is non-zero, as the tracer moves to remain on
the $u=0$ curve. In fact, the Lagrangian velocity felt by the
particle trapped on such a curve is given by
\begin{equation}
v(t)=\frac{\left. \frac{\partial u}{\p t} \right|_{u=0}}{\left.
\frac{\partial u}{\p x} \right|_{u=0}} ,
\end{equation}
with the derivatives evaluated on the curve $u=0$. For the single
scale field (\ref{usinglesc}), an explicit expression for the
Lagrangian velocity may be found in terms of the functions $f(t)$
and $g(t)$ and their derivatives:
\begin{equation}
v(t)=\frac{1}{k} \frac{f^\prime g - g^\prime f}{f^2+g^2}.
\label{veqn}
\end{equation}
When the Lagrangian velocity $v(t)$ is known, the equation of
motion for the tracer is simply
\begin{equation}
\frac{d x}{d t}=v(t), \label{Lageqn}
\end{equation}
with solution
\begin{equation}
x(t)=x(0)+\int_0^t v(t^\prime) d t^\prime.
\end{equation}
Similarly, the return on the stock over a time lag of $\Delta$ is
given by
\begin{eqnarray}
r_\Delta(t) &=& x(t+\Delta)-x(t) \nonumber\\
   &=& \int_t^{t+\Delta} v(t^\prime) d t^\prime. \label{defr}
\end{eqnarray}
Clearly the statistics of the returns follow from the statistics
of the Lagrangian velocity, indeed for short time lags it might be
expected from (\ref{defr}) that
\begin{equation}
r_\Delta(t) \approx v(t) \Delta, \label{rapp}
\end{equation}
so that the distribution of returns, for instance, can be related
directly to the distribution of the Lagrangian velocity. We shall
see later that (\ref{rapp}) is not fully accurate for large values
of $v$, but the  Lagrangian velocity
distribution will still prove extremely useful.

To find the probability distribution function $P(v)$ of the
Lagrangian velocity, we first consider the cumulative probability
that the Lagrangian velocity $v$ at a given time is greater than
some chosen value $V$. According to (\ref{veqn}), $v>V$ if and
only if
\begin{equation}
f^\prime > \frac{f^2+g^2}{g} k V+\frac{g^\prime f}{g} \equiv F,
\label{e1}
\end{equation}
where we use the symbol $F$ to denote the right-hand-side. Since
$f$, $g$, $f^\prime$ and $g^\prime$ are independent Gaussian
random variables at any single moment in time, we denote their
PDFs by $P_f$, $P_g$ etc., and the cumulative probability of the event
(\ref{e1}) is
\begin{equation}
C(V)=\int_{-\infty}^\infty dg \int_{-\infty}^\infty df
\int_{-\infty}^\infty dg^\prime \int_{F}^\infty df^\prime P_{g}
P_{f} P_{g^\prime} P_{f^\prime}.
\end{equation}
The PDF of $v$ then follows from
\begin{eqnarray}
P(v) &=& \left. - \frac{d}{d V} C(V)\right|_{V=v} \nonumber\\
  &=& \int_{-\infty}^\infty dg \int_{-\infty}^\infty df
\int_{-\infty}^\infty dg^\prime  P_{g} P_{f} P_{g^\prime} \left.
P_{f^\prime}\right|_{f^\prime=F} \left.\frac{\p F}{\p V}\right|_{V=v},
\end{eqnarray}
and the fact that
\begin{equation}
\frac{\p F}{\p V}= k \frac{f^2+g^2}{g}.
\end{equation}
Noting that the variances of $f^\prime$ and $g^\prime$ both equal $-\a^2
R^{\prime\prime}(0)=\a^2/\t_0^2$, where we introduce the notation $\t_0$ for the timescale defined by the initial radius of curvature of the function $R$:
\begin{equation}
\t_0=\frac{1}{\sqrt{-R^{\prime\prime}(0)}},\label{tau0def}
\end{equation}
the integrations over the
$f$-$g$ plane may be performed by using polar coordinates $f=r
\cos \theta$, $g=r \sin \theta$, yielding
\begin{eqnarray}
P(v) &=& \int_{-\infty}^\infty dg \int_{-\infty}^\infty df
  P_{g} P_{f} \frac{k \t_0}{\sqrt{2 \pi \a^2}} \sqrt{f^2+g^2}
  \exp\left[-(f^2+g^2)\frac{k^2 \t_0^2 v^2}{2 \a^2}\right]
  \nonumber\\
  &=& \frac{k \t_0}{2 \left[ 1+k^2 \t_0^2 v^2 \right]^\frac{3}{2}}
  \label{PDF}.
\end{eqnarray}
The PDF (\ref{PDF}) of the Lagrangian velocity $P(v)$ is symmetric
in $v$, and so has mean zero. Note that the tails of $P(v)$ decay
as $|v|^{-3}$ for large $|v|$, so the variance of $v$ does not
exist.  Also, (\ref{PDF}) effectively contains only one free
parameter $k \tau_0$, and is otherwise independent of the choice
of correlation function $R(t)$ for $f$ and $g$.

\section{Correlation of Lagrangian velocity and returns}
The (normalized) correlation of the stock returns with lag
$\Delta$, at time $t$, is defined by averaging the product of the
returns
 at times separated by $\t$:
\begin{equation}
\frac{\left<r_\Delta(t) r_\Delta(t+\t)\right>}{\left<r_\Delta(t)^2 \right>},
\end{equation}
  and for stationary
returns becomes independent of $t$ . The correlation is normalized
by the variance $\left< r_\Delta(t)^2\right>$ of the returns at
lag $\Delta$, which is known as the \emph{squared volatility} \cite{Gop}. In
this section we derive analytical results to show how the
volatility and the returns correlation for a single-scale field
with $\eps\rightarrow 0$ depend on the time correlation function
$R(t)$ of the Eulerian field, and on $\a$ and $k$.

Because of the dependence (\ref{defr}) of the returns on the
Lagrangian velocity, the correlation of the returns can be written
in terms of the correlation function of $v$:
\begin{eqnarray}
\left<r_\Delta(t) r_\Delta(t+\t)\right> &=& \int_t^{t+\Delta} d
t_1 \int_{t+\t}^{t+\t+\Delta}  d t_2 \left< v(t_1) v(t_2) \right>
\nonumber\\
&=& \int_0^\Delta dt_1 \int_\t^{\t+\Delta} dt_2 \left<
v(t_1)v(t_2)\right> \nonumber \\
&=& \int_0^\Delta dt_1 \int_\t^{\t+\Delta} dt_2 \, L(t_1-t_2)\\
&=& \int_0^\Delta  (\Delta-t_1)\left[
L(t_1-\tau)+L(t_1+\tau)\right] dt_1, \label{rcorr}
\end{eqnarray}
where we have used the stationarity of the returns, and defined
the Lagrangian velocity correlation function
\begin{equation}
L(t_1-t_2)=\left< v(t_1)v(t_2) \right>. \label{Lagcorr}
\end{equation}
Note the Lagrangian velocity is assumed to be a stationary process---this has not been  proven to follow from Eulerian stationarity in the general case \cite{Zirbel}, but appears to hold in numerical simulations.
We will proceed to calculate $L(t)$, and then use (\ref{rcorr}) to
find the volatility and returns correlation. We note here that
when the separation time $\tau$ is much larger than the lag
$\Delta$, the returns covariance (\ref{rcorr}) is approximated by
\begin{equation}
\left<r_\Delta(t) r_\Delta(t+\t)\right> \sim \Delta^2 L(\tau),
\label{rcorrLag}
\end{equation}
consistent with (\ref{rapp}).

The  correlation function of the stationary random process $v$ is
written as an eight-dimensional integral, using the definition
(\ref{veqn}) of the Lagrangian velocity:
\begin{eqnarray}
L(t) &=& \left< v(0)v(t) \right> \nonumber \\
&=& \frac{1}{k^2}\int\cdots\int \frac{f_1^\prime g_1-g_1^\prime
f_1}{f_1^2+g_1^2} \frac{f_2^\prime g_2-g_2^\prime
f_2}{f_2^2+g_2^2} P_{1 2}(f_1,f_1^\prime,f_2,f_2^\prime)
P_{1 2}(g_1,g_1^\prime,g_2,g_2^\prime). \nonumber\\\label{Lint}
\end{eqnarray}
Here the eight-fold integrals are over $f$, $g$, $f^\prime$ and
$g^\prime$ evaluated at each of the two times $t=0$ (with
subscript 1) and $t=t$ (subscript 2). The random functions $f$ and
$g$ are independent of each other, but the joint probability
functions for $f$ and its derivative at different times is found
from the Gaussian joint PDF
\begin{equation}
P_{1 2}(\mathbf{w})=\frac{1}{(2
\pi)^2}(\text{Det}\mathbf{A})^\frac{1}{2} \exp\left[-\frac{1}{2}
\mathbf{w}^T \mathbf{A} \mathbf{w} \right] \label{PGauss}
\end{equation}
for the vector of arguments,
$\mathbf{w}=(f_1,f_1^\prime,f_2,f_2^\prime)$, with a similar
expression for $g$. The matrix $\mathbf{A}$ is defined by
\begin{equation}
A_{i j}^{-1}= \left< w_i w_j\right>
\end{equation}
and when written out in full:
\begin{equation}
\mathbf{A}^{-1}= \a^2 \left\{
    \begin{array}{cccc}
    1 & 0 & R(t) & -R^\prime(t) \\
    0 & -R^{\prime\prime}(0) & R^\prime(t) & -R^{\prime\prime}(t)
    \\
    R(t) & R^\prime(t) & 1 & 0 \\
    -R^\prime(t) & -R^{\prime\prime}(t) & 0& -R^{\prime\prime}(0)
    \end{array}
    \right\} \label{Amat}.
\end{equation}
Using (\ref{PGauss}) and (\ref{Amat}) in (\ref{Lint}) enables us
to calculate the Lagrangian correlation $L(t)$. We briefly
describe here the steps in the calculation of the
multi-dimensional integrals. First, the four integrals over the
variables $f_1^\prime$, $g_1^\prime$, $f_2^\prime$, $g_2^\prime$
are performed - the dependence of the integrand on each of these
variables is linear, so the Gaussian integrals may be calculated
straightforwardly. For the remaining four variables, we transform
to polar coordinates in the $(f_1,g_1)$ and $(f_2,g_2)$ planes:
\begin{equation}
f_1=r \cos\theta, \quad g_1=r \sin \theta, \quad f_2=\rho \cos
\phi, \quad g_2 = \rho \sin \phi.
\end{equation}
The $\phi$ integral is then trivial, and by integrating over
$\theta$, $r$, and $\rho$, the final result emerges:
\begin{equation}
L(t)=\frac{1}{2 k^2 R(t)^2} \left\{ R(t) R^{\prime\prime}(t) -
R^\prime(t)^2 \right\} \log \left[ 1- R(t)^2 \right]. \label{Lres}
\end{equation}

Equation (\ref{Lres}) gives an explicit formula for the Lagrangian
velocity correlation in terms of the time correlation $R(t)$ of
the single-scale Eulerian field. When the function $R(t)$ is given, $L(t)$ follows
immediately from (\ref{Lres}), and hence the volatility and
returns correlation  can be calculated using (\ref{rcorr}). Before
examining the results for the returns, we consider the limiting
forms of $L(t)$ for small and large arguments.

Assuming that $R(t)$ is sufficiently smooth near $t=0$ to allow the definition of $\tau_0$ as in equation (\ref{tau0def}), a
small-time expansion of (\ref{Lres}) yields
\begin{equation}
L(t) \sim -\frac{1}{2 k^2 \t_0^2} \log\left(\frac{t^2}{\t_0^2}\right) \quad \text{ as }t\rightarrow
0. \label{Lsmallt}
\end{equation}
Note this diverges as $t$ approaches zero, so the
variance of the Lagrangian velocity does not exist, in accord with
our conclusion at the end of section 3.

For large time arguments, we are particularly interested in
power-law decays of correlations, and so consider how the decay of
$R(t)$ influences that of $L(t)$. Taking a power-law decay with
exponent $-\b$:
\begin{equation}
R(t) \sim t^{-\b} \quad\text{ as }t\rightarrow \infty,
\end{equation}
the corresponding Lagrangian correlation is found from
(\ref{Lres}) to decay as
\begin{equation}
L(t) \sim -\frac{\b}{2k^2} t^{-2\b-2}\quad\text{ as
}t\rightarrow\infty. \label{Llarget}
\end{equation}
The negative sign here indicates that the Lagrangian correlation
approaches zero from below at large times, being negative whenever
$R(t)$ has a power-law decay.

The squared volatility is found by setting $\tau=0$ in
(\ref{rcorr}):
\begin{equation}
\text{vol}^2(\Delta)=\left<r_\Delta(t)^2\right>= 2 \int_0^\Delta (\Delta-t) L(t) d t
\label{volres}.
\end{equation}
For short lags, with $\Delta\ll \tau_0$, it follows from
(\ref{Lsmallt}) that the volatility may be approximated by
\begin{equation}
\text{vol}^2(\Delta) \approx \frac{1}{2
k^2}\frac{\Delta^2}{\tau_0^2}\left[ 3 -
\log\left(\frac{\Delta^2}{\tau_0^2}\right)\right]+O\left(\frac{\Delta^4}{\tau_0^4}\right)
\label{volsmallt}
\end{equation}
Note that while this expression
limits to zero as $\Delta$ vanishes, it does not follow a simple
power law in $\Delta$ because of the logarithmic term. Although
empirical stock volatilities have been fitted with
power-laws \cite{Gop,Masolivier00}, we show in the next section that (\ref{volsmallt}) can
also match the data quite well.

For large time lags $\Delta$, empirical data shows a linear growth
in $\text{vol}^2$ with $\Delta$. From (\ref{volres}), we
can find an expression for the rate of change of $\text{vol}^2$
with $\Delta$:
\begin{equation}
\frac{d}{d \Delta} \text{vol}^2(\Delta) = 2 \int_0^{\Delta}
L(t)dt,
\end{equation}
and hence conclude that the squared volatility grows linearly with
$\Delta$ for large lags if the integral
\begin{equation}
\int_0^\infty L(t) dt \label{testint}
\end{equation}
has a finite positive value.
Assuming a power-law form of $R(t)$ for
large arguments, we use the result (\ref{Llarget}) for the
asymptotic form of $L(t)$. It immediately follows that the
integral (\ref{testint}) is finite if $\b>-1/2$, and hence at large lags the
squared volatility grows linearly with $\Delta$ in this case.

In empirical stock data, correlations of nonlinear functions of
the returns often display significantly slower decay than the
correlations of the returns themselves. One quantity of interest
is the correlation of absolute returns, which, using (\ref{rapp}),
may be related to the correlation of absolute Lagrangian
velocities (for short lags) as
\begin{equation}
\left< \left|r_\Delta(t)\right| \left|r_\Delta(t+\t) \right|
\right>-\left< \left|r_\Delta(t)^2\right| \right>\approx
\Delta^2\left\{ \left<\left|
v(t)v(t+\tau)\right|\right>-\left<\left|
v(t)^2\right|\right>\right\}.
\end{equation}
Following similar procedures as in the calculation of $L(t)$, it
can be shown that the large-$t$ asymptotics corresponding to a
power-law $R(t)\sim t^{-\b}$ give a correlation of absolute
Lagrangian velocities scaling as
\begin{equation}
\left< \left| v(0) v(t) ^{}\right|\right> -\left< \left| v^2
\right|\right>\sim t^{-2 \b}. \label{resabs}
\end{equation}
In contrast to (\ref{Llarget}), this correlation remains positive
at large times; note too that its rate of decay is slower than
that of (\ref{Llarget}).

Before proceeding to numerical simulations, we briefly summarize the theoretical results. We have obtained the PDF of the Lagrangian velocity in (\ref{PDF}), and related the Lagrangian velocity correlation $L(t)$ to the given Eulerian correlation $R(t)$ in (\ref{Lres}). This permits estimates of the volatility as a function of lag. Finally, we expect a power-law scaling for the absolute returns correlation when $R(t)$ has a power-law decay. In the following sections we perform numerical simulations to confirm and extend these results.  

\section{Numerical simulations}
To extend the analytical results found in the previous section, we
consider the implementation of  numerical simulations of motion in
random fields. Random fields $u(x,t)$ may be generated using
standard techniques \cite{MajdaKramer}, with the ordinary
differential equation (\ref{6red}) being solved using Runge-Kutta
methods. Averages may be calculated over an ensemble of
realizations, or over a long time series $x(t)$ in a single
realization, to closely mimic the statistical analysis of S\&P 500
data performed in \cite{Nat95, Gop}.

A random field with zero mean and correlation function (\ref{8red})
may be generated using a superposition of random Fourier modes \cite{MajdaKramer}:
\begin{equation}
u(x,t)=\frac{1}{\sqrt{N}}\sum_{n=1}^N A_n \cos(\w_n t+k_n x)+B_n
\sin(\w_n t+k_n x), \label{num1}
\end{equation}
with the amplitudes $A_n$ and $B_n$ chosen from independent
Gaussian distributions of zero mean and variance $\a^2$. The
$\w_n$ and $k_n$ are chosen from distributions of random numbers
chosen so as yield the correlation (\ref{8red}). Specifically, the
the $\w_n$ are chosen from a distribution shaped as the Fourier
transform of $R(t)$, with the distribution of the $k_n$ being the
Fourier transform of $S(x)$. Thus, Gaussian distributions with
zero mean and unit variance are used for the $\w_n$ and $k_n$
to generate the random field of Fig. 1, as the Fourier
transforms yield $R(t)=\exp(-t^2/2)$ and $S(x)=\exp(-x^2/2)$ as
required in equations (\ref{Rexp}) and (\ref{Sfig2}).

For the special case of a single-scale field considered here, the
$k_n$ are all  $\pm k$, and the field may be written in
the simpler form  (\ref{usinglesc}), see \cite{exact}. Accordingly
we only require a method for generating the random functions of
time $f(t)$ and $g(t)$, and this is easily derived from
(\ref{num1}):
\begin{equation}
f(t)=\frac{1}{\sqrt{N}} \sum_{n=1}^N A_n \cos(\w_n t)+B_n
\sin(\w_n t), \label{num2}
\end{equation}
with a similar formula for $g(t)$. The $A_n$, $B_n$ are chosen as
above. We are especially interested in the effects of power-law
correlations $R(t)$, so we choose the $\w_n$ in (\ref{num2}) from
the Gamma distribution \cite{MajdaKramer}:
\begin{equation}
G(\w)=\frac{T^\b}{\Gamma(\b)}\w^{\b-1}e^{-T \w}.
\end{equation}
Here $\Gamma(\b)$ denotes the usual Gamma function. The Fourier
transform of $G(\w)$ is the correlation function $R(t)$ of $f$:
\begin{equation}
R(t)=\left(1+\frac{t^2}{T^2}\right)^{-\frac{\b}{2}}\cos\left[ \b
\tan^{-1}\left(\frac{t}{T}\right)\right], \label{Rpow}
\end{equation}
which decays as $R(t)\sim t^{-\b}$ for $t\gg T$.

The methods described above are sufficient to simulate motion in a
random field as described by (\ref{6red}), or equivalently by the
rescaled equation (\ref{5resc}), and indeed this is how the stock log curves in Figs. 1 to 3 are generated. However, in the idealized limit
$\eps \rightarrow 0$ considered in the previous sections, the tracer closely follows the contour $u=0$.
In this limit it is therefore not necessary to explicitly solve
the differential equation to determine $x(t)$ in a single-scale field, as this can be
determined from the implicit equation $u(x,t)=0$. From
(\ref{usinglesc}) we immediately obtain
\begin{equation}
x(t)=\frac{1}{k}\tan^{-1}\left[ -\frac{f(t)}{g(t)} \right] \label{num3}
\end{equation}
for motion along the $u=0$ contour. Given the expression
(\ref{num2}) for $f(t)$ and $g(t)$, equation (\ref{num3}) then
generates the time series $x(t)$. Some care must be taken to
ensure continuity of $x(t)$ near times when $g(t)=0$, but overall
this method is a very efficient way to create a model time series
for $x(t)$.

\section{Numerical results}
In this section we report the results of employing the algorithm
(\ref{num3}) to generate a sample time series $x(t)$ corresponding
to a single-scale field, with Eulerian time correlation function
given by (\ref{Rpow}).  Having decided upon a single-scale field
and the $\eps\rightarrow 0$ limit, so that the stock tracer
follows contours of $u(x,t)=0$, our remaining choices to fit
empirical stock data is rather limited. We are free to choose the
time correlation function $R(t)$ and the spatial scaling factor
$k$; note the standard deviation $\a$ of the random field has
disappeared because we have take the limit $\eps\rightarrow 0$.

Our choice for $R(t)$ is motivated by the empirical results such
as those reported in \cite{Gop}, and by our analysis in previous
sections. We have seen in section 4 that the Lagrangian velocity correlation
becomes negative, with a power-law decay exponent of $-2\b-2$ when
$R(t)$ has a power-law decay exponent of $-\b$. On the other hand,
the correlation of absolute values of the velocity remains
positive for large $t$, with a decay exponent $-2\b$. Studies of
the corresponding returns correlations in \cite{Gop} have
concluded that the correlation decays exponentially quickly to the
noise level, whereas the absolute returns correlation exhibits a
slow decay with power-law exponent of approximately $-1/3$.
Matching this to our result (\ref{resabs}) motivates us to choose
a power-law form for $R(t)$ as given in (\ref{Rpow}), taking the
value of $\b=1/6$. The time-scaling $T$ is chosen to be 10
minutes, to approximately match the volatility results of \cite{Gop}. Having
thus defined the function $R(t)$, we are left with only the single
parameter $k$. This is chosen to be 500, to approximately fit the
volatility results (see below) to those in \cite{Gop}.

Each realization consists of $1.6\times10^6$ values of $x(t)$,
modelling the log price at 1-minute intervals, to mimic the
S\&P500 data set used in \cite{Nat95}. Averaging is over time
within each realization; we show results from different
realizations only when statistical scatter is evident in the
single-realization results. From the series for $x(t)$, it is
straightforward to calculate the  series of returns over integer
time lags $\Delta$
\begin{equation}
 r_\Delta(t) = x(t+\Delta)-x(t) .
 \end{equation}
 and to calculate various statistical properties of the returns.

The volatility
\begin{equation}
 \text{vol}(\Delta) =\sqrt{
\left< r_\Delta(t)^2 \right>}
\end{equation}
found by numerical simulation is plotted as a function of the lag
$\Delta$ in Fig. 4 (filled circles). The parameter $k$ is chosen
as $k=500$, in order to closely match the empirical $\Delta=1$ volatility value as
shown in Fig. 3(c) of \cite{Gop}. Also shown in Fig. 4 is the
analytical result (\ref{volres}), plotted as a dotted line, which
confirms that the numerical method closely reproduces the exact
results of section 4. The dashed line in Fig. 4 is the
small-lag approximation (\ref{volsmallt}), and shows that the
super-diffusive region of volatility (for lags under 10 minutes)
is well-matched by the logarithmic-corrected quadratic power law
in $\Delta$. Analysis of the empirical data frequently leads to
fitting the super-diffusive volatility with a power-law, with
exponents in the range 0.67 \cite{Gop} to 0.77
\cite{Masolivier00}. At longer lags ($\Delta>10$), random walk
behavior ($\text{vol}\sim\sqrt{\Delta}$) is observed in all
studies, and reproduced in our model as shown following equation
(\ref{testint}). The solid lines in Fig. 4 show power-law
scalings, with exponents 0.77 (for $\Delta<10$), and 0.5 (for
$\Delta \gg 10$), which are found in \cite{Masolivier00} to match the
S\&P500 volatility. Clearly these scalings also match the results
of our model quite well, indicating that the model predictions
behave similarly to the empirical volatility curve. Indeed, our result
(\ref{volsmallt}) predicts that a log-corrected quadratic fit
should be superior to the power-law fits used in \cite{Gop,
Masolivier00}. Testing this prediction requires higher frequency
analysis of the empirical market data.

The probability distribution function (PDF) of the one-minute
returns, $r_1(t)$, is plotted in Fig. 5. Open circles are from
numerical simulations, and are well-fitted by the Lagrangian
velocity PDF (\ref{PDF}), up to high returns values. Note the
returns are normalized by their standard deviation $\sigma\equiv
\text{vol}(1)=2\times 10^{-4} $. The
Gaussian distribution with this standard deviation is plotted with
the dotted line: note the higher-than-Gaussian central peak of the
model results, and the fatter tails. Fig. 5 is remarkably similar to
Fig. 2 of \cite{Nat95}, which shows the one-minute returns PDF
from the S\&P500. Indeed, in \cite{Nat95} it is demonstrated that
the empirical results are well-fitted near the center of the
distribution by a L\'{e}vy stable distribution with index
$\a=1.4$, and so in Fig. 5 we also show this distribution (dashed
line), with scale factor chosen to match the peak value of the
numerical distribution. Comparison with Fig. 2 of \cite{Nat95}
shows that our numerical simulation returns have a very similar
distribution to the S\&P500 data, at least within $\pm 10 \sigma$.

In Fig. 6 we consider the correlation function of the one-minute
returns over a separation time $\tau$, normalized by the
volatility:
\begin{equation}
\frac{\left< r_1(t)r_1(t+\t)\right>}{\left< r_1(t)^2\right>}.
\label{numretcorr}
\end{equation}
As this quantity has both positive and (small) negative values, we
plot its absolute value on a log-linear scale. Statistical scatter
causes some uncertainty in these numerical results, so the
results of two different realizations (circles and triangles,
respectively) are shown. According to (\ref{rcorrLag}) and (\ref{Llarget}),
the returns correlation is negative for $\tau\gg 10$ minutes,
which is confirmed by plotting the analytical result
(\ref{rcorr}): this is shown as a solid line where the correlation
is positive, and a dashed line where it is negative. The numerical
simulation results match the analytical form well for $\tau<10$
minutes, but degrade in quality when the correlation is negative. It is possible that more advanced numerical simulation methods such as Fourier-Wavelet \cite{MajdaKramer} could alleviate this problem. 
Note filled symbols show positive numerical correlation values,
with open symbols for negative values. The rapid decay of the
correlation from $\tau=0$ to $\tau=10$ minutes (see dotted line showing $\exp(-\tau/3)$ for comparison) is reminiscent of
that found in S\&P500 data, see Fig. 3(a) of \cite{Gop} and Fig. 8(a) of \cite{Liu}. Unlike our model results, however, the empirical correlation does not exhibit
negative values but instead reaches a ``noise level''
around $3\times 10^{-3}$, where it flattens out without
discernable structure. Our model suggests that it might be
fruitful to search for evidence of negative correlations in the
empirical data, which may currently be screened by the noise
level.

The correlation function of the absolute value of one-minute
returns (normalized to unity at zero separation) is
\begin{equation}
\frac{
\left<\left|r_1(t)r_1(t+\tau)\right|\right>-\left<\left|r_1(t)\right|\right>^2}
{\left<\left|r_1(t)^2\right|\right>-\left<\left|r_1(t)\right|\right>^2}
\label{numabscorr}.
\end{equation}
From empirical data, this is known to have a slow power-law decay
with $\tau$, with exponent of approximately $-1/3$, see Fig. 3(b)
of \cite{Gop} and Fig. 8(b) of \cite{Liu}. The analytical result
(\ref{resabs}) for the present model indicates that a similar
power-law form holds, indeed this is what motivated our choice of
parameter $\beta=1/6$ in (\ref{Rpow}). Fig. 7 shows numerical
results from two separate realizations (circles and triangles),
with the solid line showing a power-law decay with exponent
$-1/3$. The numerical correlations are all positive, but show a
slight deviation from the expected power-law scaling. This may be
related to the poor representation of the negative values of the
returns correlation in Fig. 6. Comparison with empirical data
(Fig. 3(b) of \cite{Gop} and Fig. 8 (b) of \cite{Liu}) also shows
that the model's correlation decays faster than that of the
S\&P500 data for $\tau<10$ minutes.

The non-Gaussian distribution of returns over longer time lags is
examined in Figs. 8 and 9, using the methods applied to S\&P500
data in Figs. 6 to 8 of \cite{Gop}. The normalized returns are defined by dividing by the volatility:
\begin{equation}
g_\Delta(t)=\frac{r_\Delta(t)}{\text{vol}(\Delta)},
\label{normret}.
\end{equation}
and the moments
\begin{equation}
\left< \left|g_\Delta(t)\right|^m\right> \label{defmom}
\end{equation}
 of the numerical simulations are
plotted in Fig. 8(a) as a function of $m$, for lags of 1, 16, 32,
64, and 128 minutes. The extremely non-Gaussian distribution at
$\Delta=1$ minute relaxes  to a shape which remains stable up to
$\Delta\approx 128$ minutes---note the moments for lags from 16 to
128 minutes are almost indistinguishable in Fig. 8(a). This
conclusion is supported by the cumulative distribution of the
normalized returns shown in Fig. 8(b). This is very similar to the
behavior observed in empirical data, see Fig. 6 of \cite{Gop},
although the model does not produce power-law scaling of the
distribution tails (except for the region with exponent -2 for
lag $\Delta=1$ minute). The returns  distributions eventually
converge to a Gaussian distribution, see the corresponding plots
for lags of 390, 780, and 1560 minutes (1 to 4 trading days) in
Fig. 9. This convergence to Gaussian is somewhat faster that that
observed in the S\&P500 in \cite{Gop}, where non-Gaussian moments
and cumulative distributions persist until approximately 4 days.
Our corresponding estimate for the model persistence time (based
on the moments in Fig. 8(a)) is 128 minutes, an order of magnitude
smaller that the S\&P500 value.

The scaling of the peak $P(0)$ of the PDF of \emph{stock price
returns} $S(t+\Delta)-S(t)$ was examined in \cite{Nat95} for
S\&P500 data. A power-law scaling of $P(0)$ with lag was
discovered, with exponent $-0.7$. In Fig. 10 we mimic this study,
generating the time series $S(t)$ from $x(t)$ in order to
calculate the stock price return. Results are shown from two
separate realizations, along with a line indicating Gaussian
scaling of exponent $-1/2$. It is clear that, unlike the S\&P500
returns, the model $P(0)$ does not scale as a single power-law
with $\Delta$, although it appears to approach the Gaussian
scaling at the highest lags.

\section{Conclusion}
We have proposed a new model for the fluctuations of stock prices,
in which the rate of change of the log price is randomly dependent
upon both time and the current price, see equation (\ref{6red}).
This is a very general modelling concept, with the attractive
feature that Gaussian Eulerian fields may give non-Gaussian
Lagrangian statistics in the  observable data, i.e. the time
series of returns. As noted in section 5, Monte-Carlo simulations
of equation (\ref{6red}) can be used to find the model predictions
for any random field. In this paper we focus on a special case,
the $\epsilon\rightarrow 0$ limit of a single-scale random field,
for which exact analytical results are obtainable, and for which
numerical simulations are particularly efficient.

Our main analytical results are the expression (\ref{Lres}) for the
Lagrangian velocity correlation $L(t)$ in terms of the given
Eulerian field correlation $R(t)$, and the quadrature formulas
(\ref{rcorr}) and (\ref{volres}) giving the correlation function
and volatility of the returns. Numerical simulations of stock
price time series are efficiently performed using equation
(\ref{num3}), and allow us to examine features not amenable to
exact analysis.

We find that several of the important empirical stylized facts
(i)-(iv) listed in the Introduction are reproduced by our model,
using the correlation function (\ref{Rpow}) for computational
convenience. The parameter values $k$ and $T$ are chosen by
comparing the model's volatility to the data in Fig. 3(c) of
\cite{Gop}. With reference to the stylized facts listed in section
1, the results of the single-scale random field model may be
summarized as follows.
\begin{itemize}
\item[(i)] Model returns are non-Gaussian, with fat tails and the
center of the distribution well-fitted by the L\'{e}vy
distribution used in \cite{Nat95}. The tails of the cumulative distributions of model returns
do not, however, have the power-law scaling found in \cite{Gop}. The normalized model returns distributions
exhibit a slow return toward Gaussian, on timescales an order of
magnitude longer than the characteristic time of $R(t)$. However,
these timescales are an order of magnitude smaller those found for
the convergence to Gaussian in the S\&P500 returns in  \cite{Gop}.

 \item[(ii)] The correlation function of model returns decays
 quickly (similar to exponentially) over a short timescale.
 Following the fast decay, the correlation becomes negative, with
 magnitude decaying more slowly to zero. Although the fast decay
 for lags under 10 minutes is very similar that found for the
 S\&P500 in \cite{Gop},
 the empirical correlation function does not reach negative
 values, rather remaining at a constant ``noise level''. Whether this noise level could be
 masking negative correlation values as predicted by the model is a
 question requiring further extensive analysis of market
 data.

\item[(iii)] The volatility of the model returns grows diffusively
for lag times longer than 10 minutes. The super-diffusive
volatility at higher frequencies (shorter lags) is actually due to a logarithmic
correction to power-law growth (\ref{volsmallt}), but for the range of lags
examined in \cite{Gop, Masolivier00} it matches well to a
super-diffusive power-law. Higher frequency data is needed to
determine which form best fits the actual stock market volatility.

\item[(iv)] Our model does not have a simple scaling of $P(0)$
with lag, in contrast to the case of S\&P500 prices
as demonstrated in Fig. 1 of \cite{Nat95}.
\end{itemize}
In summary, the single-scale random field model yields time series
which have many (but not all) of the important properties of
empirical data. The model gives an appealingly intuitive picture
of the causes of non-Gaussian behavior in markets, and provides a
simple algorithm for generating time series with many market-like
features. We anticipate that this algorithm will prove
particularly useful in Monte-Carlo simulations to determine prices
of options and derivatives when non-Gaussian returns are
particularly important \cite{Hull, Campbell}.

The concept of modelling stock prices by motion in a random field
is quite general, and suggests many directions for further work.
Among these are:
\begin{itemize}
\item[(a)] Examining the consequences of such models for the
pricing of options and derivatives, and comparison with the
standard Black-Scholes (random-walk-based) results.

\item[(b)] Further market data analysis, to determine if there is
any evidence of the negative correlation or log-corrected
volatility scaling discussed in (ii) and (iii) above.

\item[(c)] Finding a microeconomic basis for the random field model (\ref{6red}), possibly based on the actions of multiple independent agents \cite{Lux, Appleby}.

\item[(d)] Numerical simulations of (\ref{6red}) for more general
field correlations $S(x)$ and different $\eps$ values, to
determine which of the results detailed here are universal
features of motion in random fields, and which are specific to the
single-scale, $\eps\rightarrow 0$ case analyzed here.
\end{itemize}
\section{Acknowledgements}
The author gratefully acknowledges helpful discussions with John Appleby and Paul O'Gorman. This work is supported by funding from a Science Foundation Ireland Investigator Award 02/IN.1/IM062, and from the Faculty of Arts Research Fund, University College Cork.

\clearpage

\begin{figure}
\centering \epsfig{figure=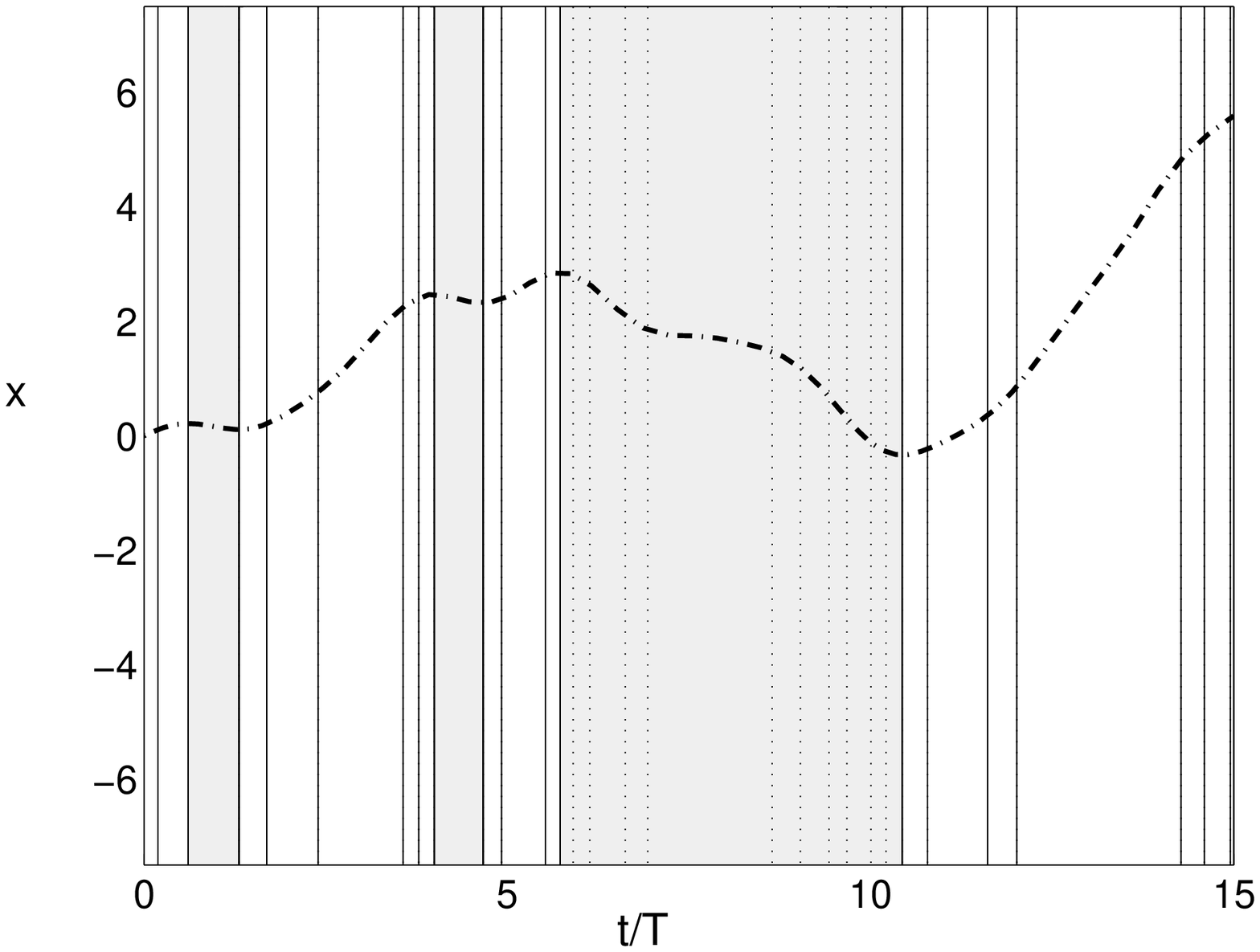,height=2.6in}
\caption{Motion of a stock tracer in a random field which depends only on time, as in (\ref{4red}). Shaded areas and dotted contour lines indicate negative values of $u_c(t)$.} \label{figtnoise}
\end{figure}
\begin{figure}
\centering \epsfig{figure=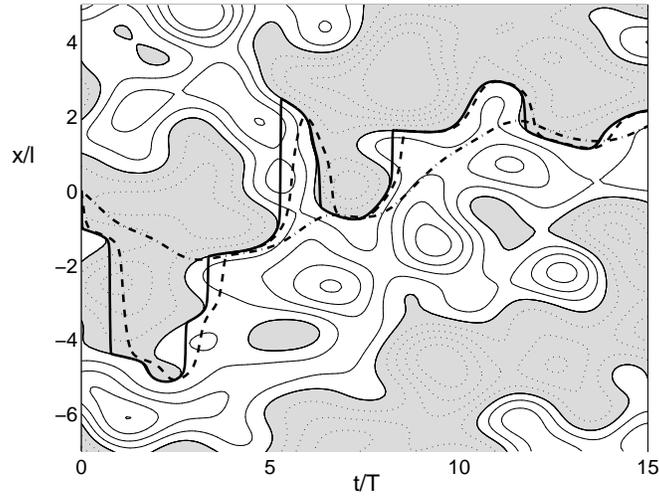,height=2.6in}
\caption{Motion of stock tracers in a random field which depends on both time and $x$, as in (\ref{6red}). The correlation functions are (\ref{Rexp}) and (\ref{Sfig2}), and the values of $\eps$ are: $\eps=1.0$ (dash-dotted), $\eps=0.1$ (dashed), and $\eps=0.01$ (solid).} \label{figcont1}
\end{figure}
\clearpage
\begin{figure}
\centering \epsfig{figure=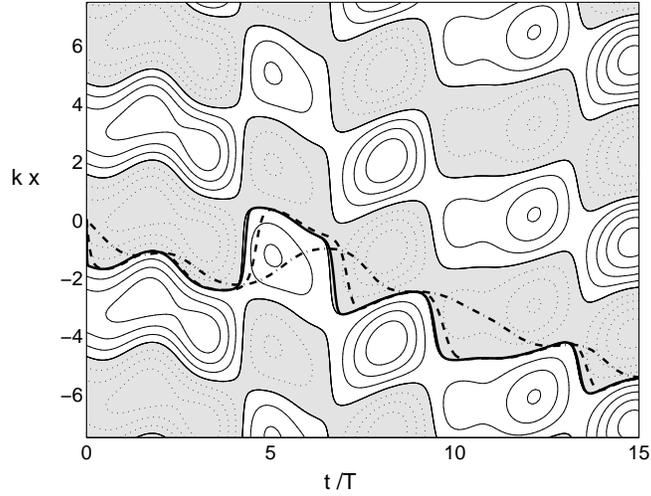,height=2.6in}
\caption{Motion of stock tracers in a single-scale random field. See Fig. \ref{figcont1} for description of line types.} \label{figss}
\end{figure}
\begin{figure}
\centering \epsfig{figure=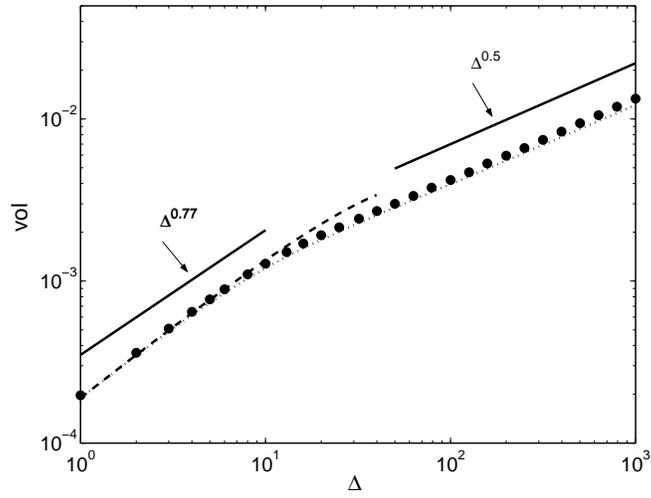,height=2.6in}
\caption{Volatility of returns as a function of lag time $\Delta$. Symbols show the results of numerical simulation; dotted line is the analytical expression (\ref{volres}); dashed line is the small-$\Delta$ approximation (\ref{volsmallt}); solid lines show power-laws with exponent $0.5$ (large lags), and $0.77$ (small lags), as fitted to empirical data in \cite{Masolivier00}.} \label{figvol}
\end{figure}
\begin{figure}
\centering \epsfig{figure=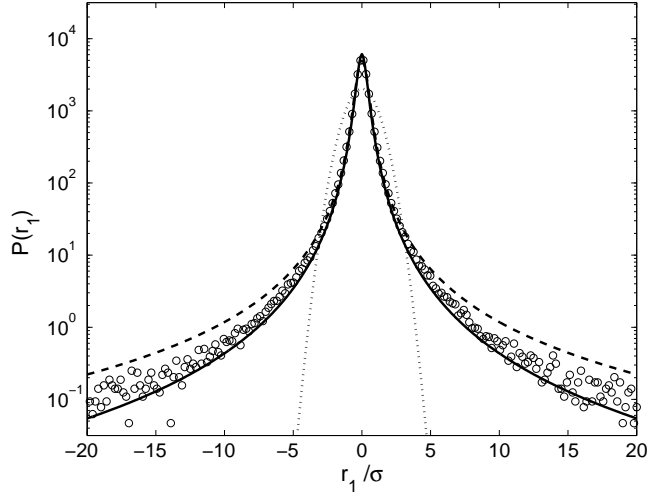,height=2.6in}
\caption{Probability distribution function of 1-minute returns from numerical simulation (symbols). Dashed line is the L\'{e}vy distribution fitted to empirical data in \cite{Nat95}; solid line is the Lagrangian velocity PDF (\ref{PDF}); dashed line is a Gaussian distribution with same variance as the data. } \label{figPDF}
\end{figure}
\begin{figure}
\centering \epsfig{figure=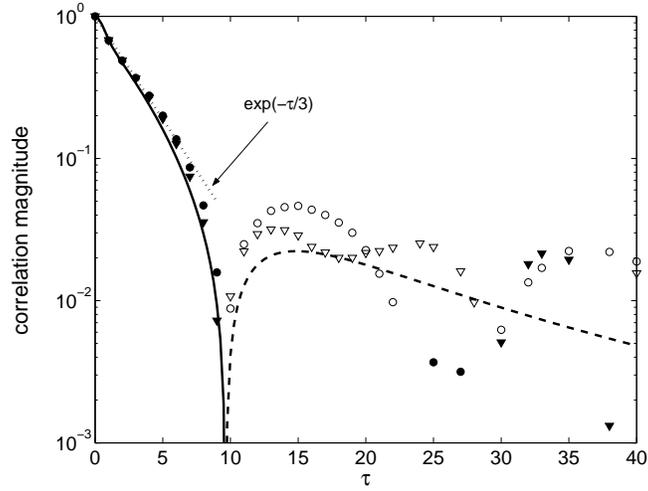,height=2.6in}
\caption{Magnitude of the correlation function of 1-minute returns, as a function of separation time $\tau$. Numerical results from two realizations are shown (circles and triangles), with the analytical form (\ref{rcorr}) (solid and dashed lines). Negative correlation values are shown with open symbols and dashed line. The dotted line shows $\exp(-\tau/3)$ for comparison.} \label{figcorr}
\end{figure}
\begin{figure}
\centering \epsfig{figure=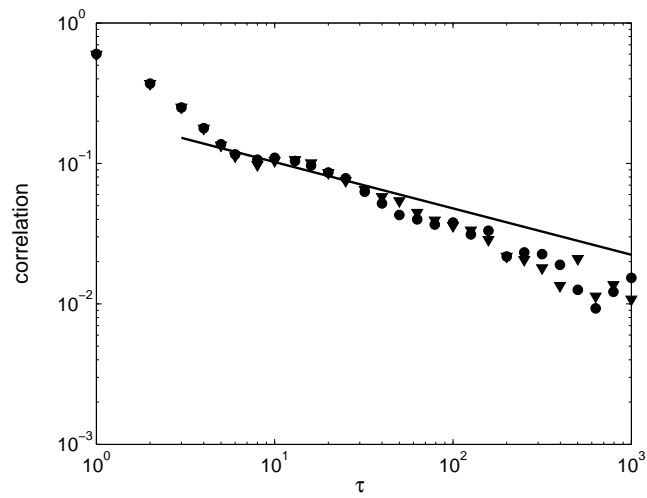,height=2.6in}
\caption{Correlation function of absolute 1-minute returns, as a function of separation time $\tau$. Numerical results from two realizations are shown (circles and triangles). The solid line is a power-law scaling as (\ref{resabs}).} \label{figabscorr}
\end{figure}
\begin{figure}
\centering \epsfig{figure=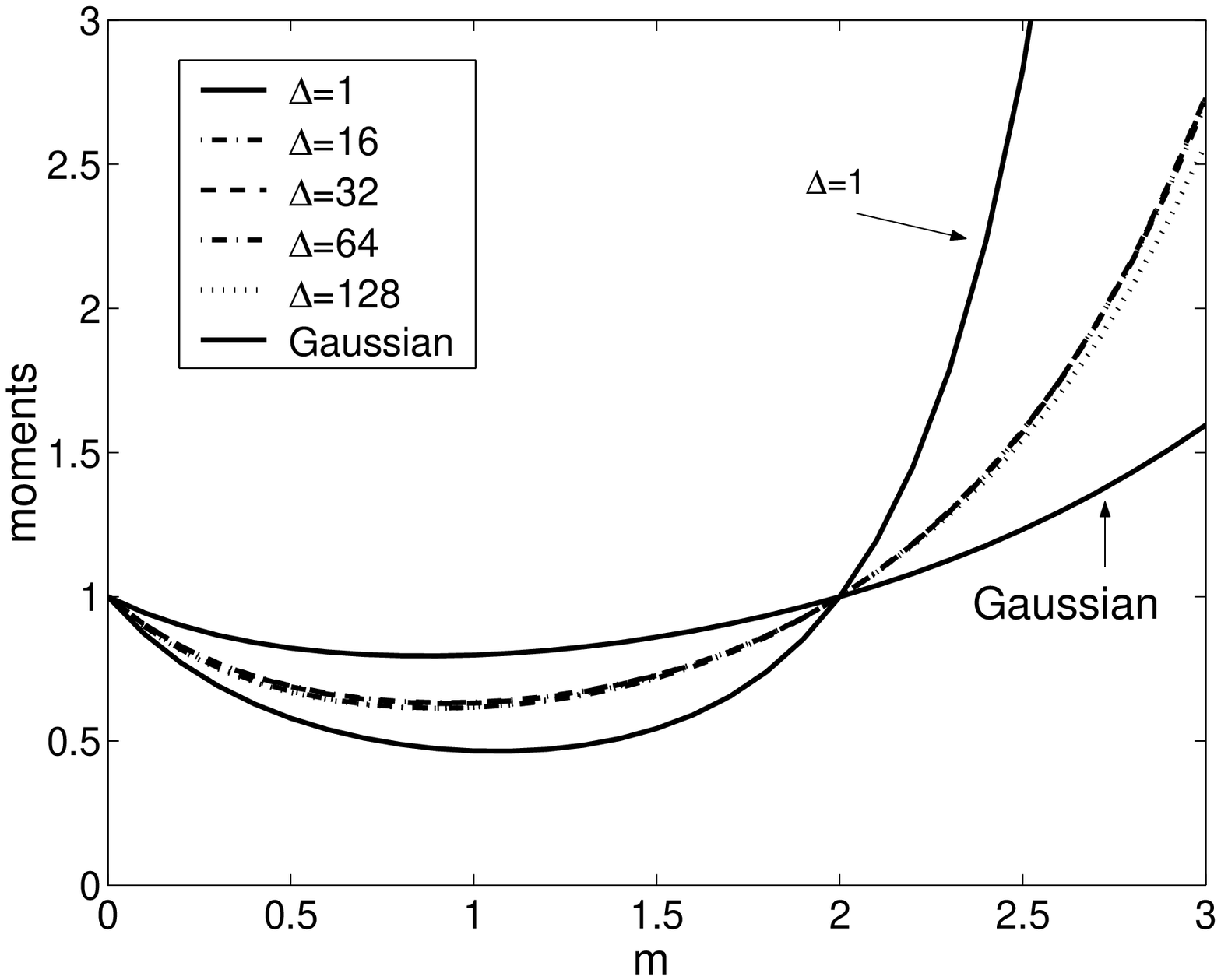,height=2.6in}
\epsfig{figure=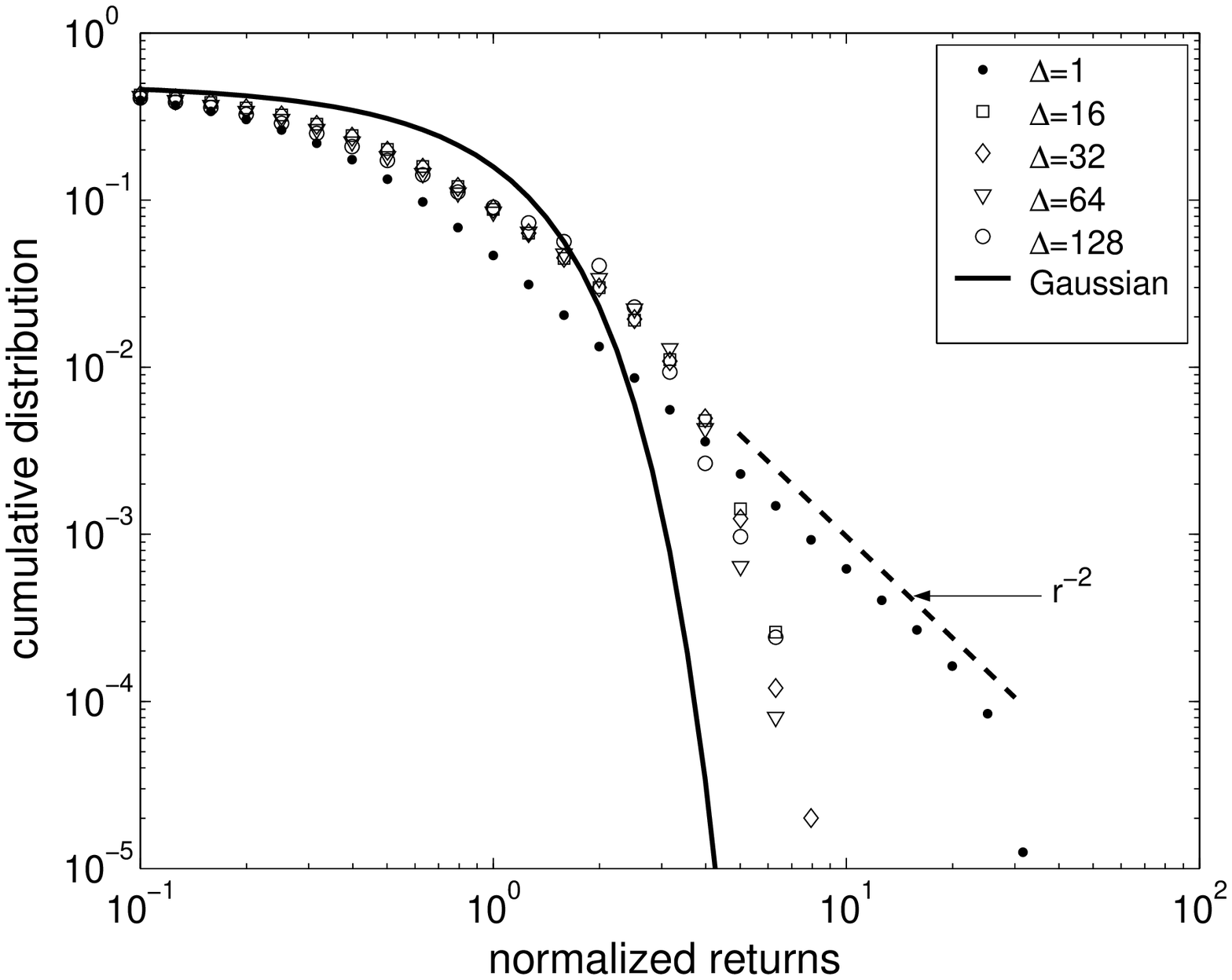,height=2.6in}
\caption{(a) Moments of the normalized returns, for lag times $\Delta=1$, 16, 32, 64, and 128 minutes. (b) Corresponding cumulative distribution functions, with the dashed line showing a power-law with exponent -2. } \label{figcumAab}
\end{figure}
\begin{figure}
\centering \epsfig{figure=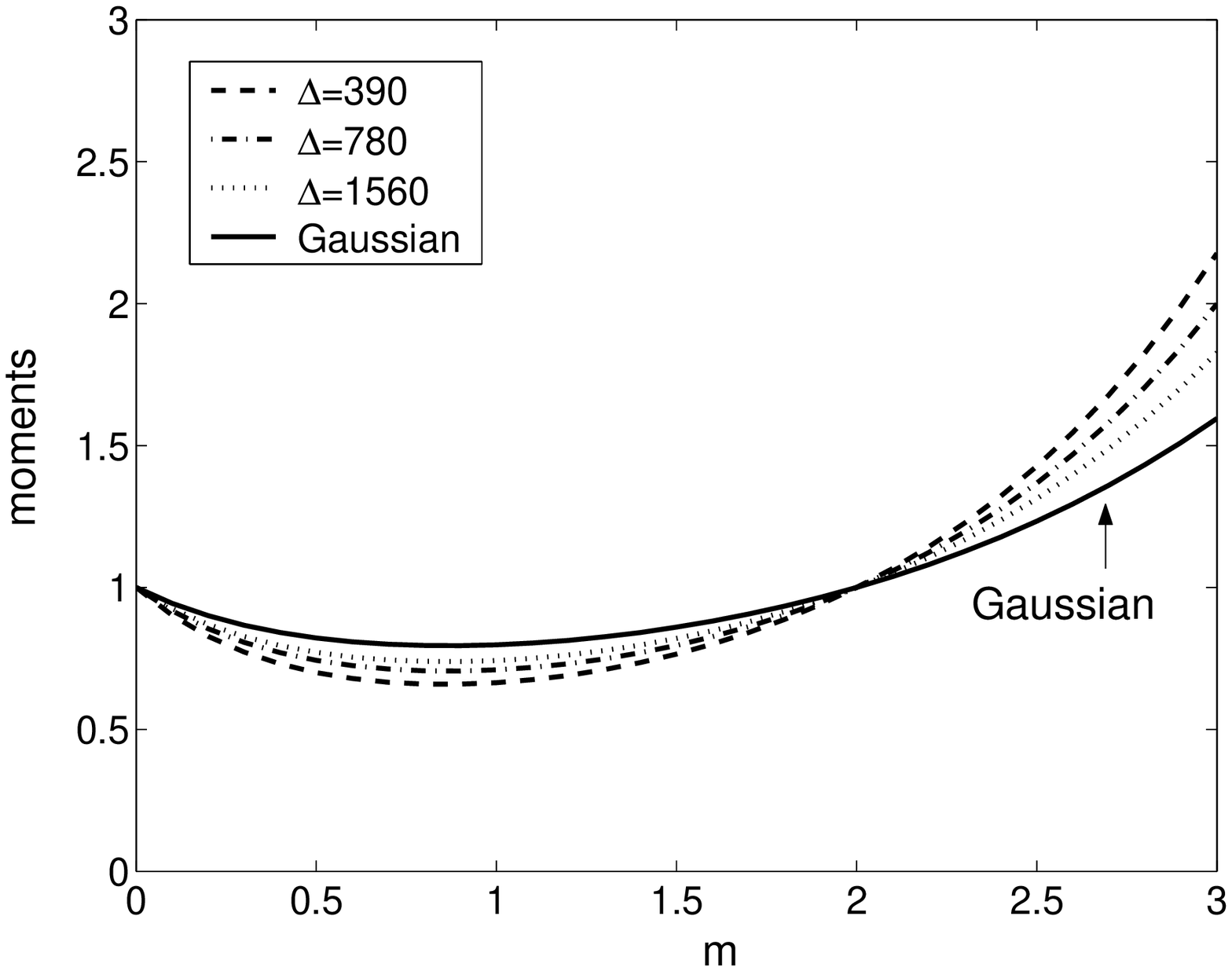,height=2.6in}
\epsfig{figure=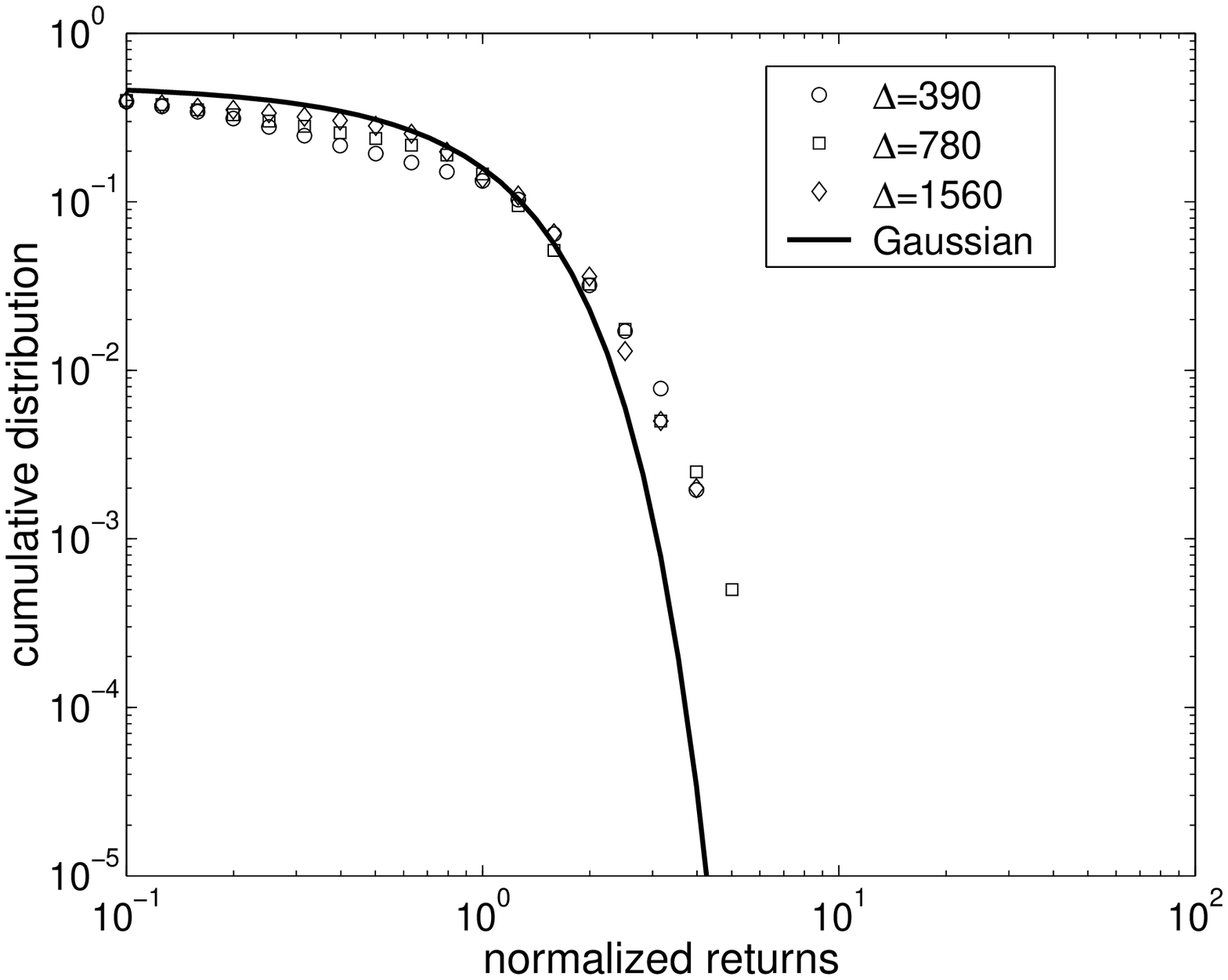,height=2.6in}
\caption{(a) Moments of the normalized returns, for lag times $\Delta=390$, 780, and 1560 minutes. (b) Corresponding cumulative distribution functions.} \label{figcumBa}
\end{figure}
\begin{figure}
\centering \epsfig{figure=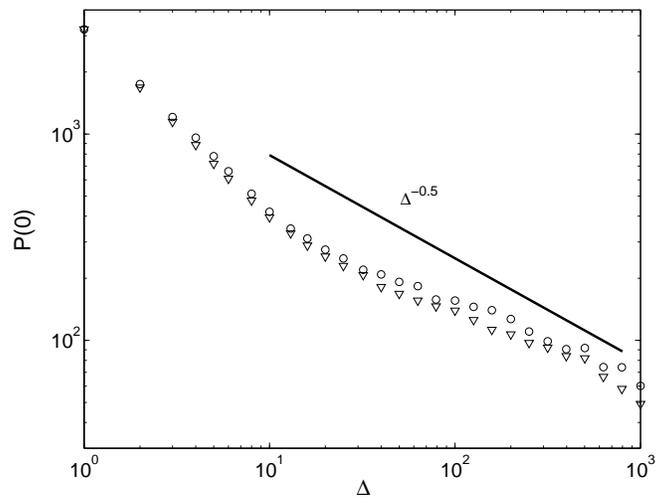,height=2.6in}
\caption{Scaling of the peak $P(0)$ of the stock returns PDF with lag time $\Delta$. Solid line shows a Gaussian scaling, with exponent $-0.5$. } \label{figP0scaling}
\end{figure}
\end{document}